\documentclass[aps,prd,twocolumn,amsmath,amssymb,amsfont,superscriptaddress,nofootinbib]{revtex4-1}
\usepackage{amsmath,aas_macros,xcolor,graphicx}
\newcommand{\bs}{\boldsymbol}
\newcommand{\diff}{{\mathrm d}}

\newcommand{\T}{\bold{T}}
\newcommand{\I}{\bold{I}}
\newcommand{\spin}{\bs{j}}
\newcommand{\Spin}{\bs{L}}

\begin{document}
\title{Spin Mode Reconstruction in Lagrangian Space}

\author{Qiaoya Wu}
\affiliation{Department of Astronomy, Xiamen University, Xiamen, Fujian 361005, China}
\author{Hao-Ran Yu}\email{\url{haoran@xmu.edu.cn}}
\affiliation{Department of Astronomy, Xiamen University, Xiamen, Fujian 361005, China}
\author{Shihong Liao}
\affiliation{Department of Physics, University of Helsinki, Gustaf Hällströmin katu 2, FI-00560 Helsinki, Finland}
\author{Min Du}
\affiliation{Kavli Institute for Astronomy and Astrophysics, Peking University, Beijing 100871, China}
\date{\today}

\begin{abstract}
Galaxy angular momentum directions (spins) are observable, well described
by the Lagrangian tidal torque theory, and proposed to probe the primordial
universe.
They trace the spins of dark matter halos, and are 
indicators of protohalos properties in Lagrangian space.
We define a Lagrangian spin parameter and tidal twist parameters 
and quantify their influence on the spin conservation and predictability 
in the spin mode reconstruction in $N$-body simulations.
We conclude that protohalos in more tidal twisting environments are
preferentially more rotation-supported, and more likely to conserve their 
spin direction through the cosmic evolution.
These tidal environments and spin magnitudes are predictable by
a density reconstruction in Lagrangian space,
and such predictions can improve the correlation between galaxy spins
and the initial conditions in the study of constraining the primordial
universe by spin mode reconstruction.

\end{abstract}

\maketitle

\section{Introduction}\label{sec.intro}
The large scale structure (LSS) of the Universe contains plenty of
cosmological information.
One of the most important tasks of LSS studies is to use the distribution
of galaxies to interpret the initial conditions of the Universe
and cosmological parameters.
The standard way is to relate the number density of galaxies in redshift
space to the primordial density perturbation in Lagrangian space,
including complications of survey geometries, biases of tracers and selection functions,
redshift space distortion (RSD), nonlinear evolution and loss of cosmological
information, etc.

\begin{figure*}
	\centering
	\includegraphics[width=0.8\linewidth]{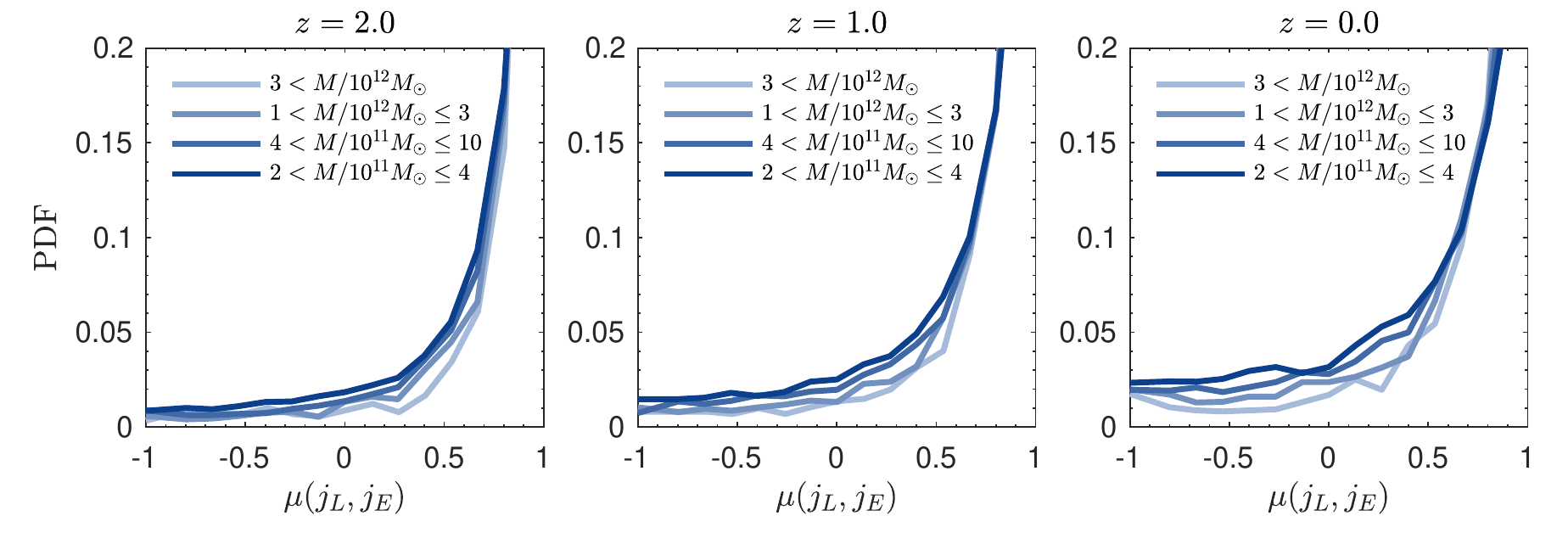}
	\caption{Conservation of the spin directions of dark matter halos 
	shown by the probability distribution functions (PDFs) 
	of $\mu(\spin_L,\spin_E)$ measured in different redshifts. 
	The case of $\mu(\spin_L,\spin_E)=1$ corresponds to a fact that
	a halo and its protohalo have exact same spin directions, where
	the spin direction is maximumly conserved.
	$\mu$ obeys a tophat distribution between $-1$ and $1$ 
	for uncorrelated vectors.
	In the three panels, the PDFs obviously departure from 
	tophat distributions and suggest the conservation of halo spin.
	Curves with different colors represent 
	the PDFs of $\mu(\spin_L,\spin_E)$ in different mass groups.}
	\label{fig.mu_z}
\end{figure*}

The rotation of galaxies adds another degree of freedom to probe the
$B$-mode clustering of matter, and 
is proposed to provide additional information of the primordial universe.
The tidal torque theory elucidates the basic mathematical framework that
the initial misalignment between the moment of inertia of protohalos 
and the tidal fields they feel 
can produce persistent tidal torque 
\cite{1969ApJ...155..393P,1970Afz.....6..581D,1984ApJ...286...38W}.
In SCDM (standard cold dark matter cosmology, 
a cosmological model without the cosmological constant,  
$\Omega_m=1, \Omega_\Lambda=0$)
$N$-body simulations, 
\cite{2000ApJ...532L...5L} first quantified 
that this tidal torque is strongly correlated with the final halo spin
and proposed using galaxy spins to reconstruct the initial tidal field
\cite{2001ApJ...555..106L}.
Although the spins of dark matter halos are not directly observable,
hydrodynamic galaxy formation simulations all suggest that
the spins of central galaxies and the spins of their host halos are
highly correlated (e.g. \cite{2015ApJ...812...29T} and references therein).
In Lagrangian space, by using the $E$-mode reconstructed density fields
\cite{2017PhRvD..95d3501Y}
we are able to construct a scale dependent spin mode,
and successfully predict the spins of dark matter halos 
\cite{2020PhRvL.124j1302Y}.
By applying this method to the ELUCID reconstruction of the local
universe \cite{2016ApJ...831..164W} and a first
spin catalog of observed galaxies, \cite{2020NatAs.tmp..241M} for the
first time confirmed the correlation between galaxy spins and initial 
conditions of the Universe.
These studies open the path of using galaxy spins to constrain the 
primordial state of the universe, including primordial non-Gaussianity,
gravity waves, parity violation \cite{2020PhRvL.124j1302Y}, 
and neutrino mass \cite{2019PhRvD..99l3532Y}.
 
In order to maximize the constraining power, an important task is to
understand galaxy spin - initial condition correlation as a function
of many possible aspects: nonlinear evolution, quality of
$E$-mode reconstruction, and galaxy types and their Lagrangian space properties.
Besides the spin directions, the magnitude of the angular momentum vector has 
been extensively studied in many literatures, 
usually parameterized by spin parameters.
The spin parameters characterize the rotation-supportedness of the dark matter
halos in Eulerian space, and can not be directly measured since matter distribution 
and velocity dispersion of dark matter cannot be acquired 
simply through galaxy spectroscopy. A number of studies show that
the morphologies of galaxies and galactic spin parameters
are correlated \cite{1983IAUS..100..391F,2012ApJS..203...17R}
(disk galaxies are rotation-supported as assumed in the 
classical disk galaxy formation model \cite{1980MNRAS.193..189F,1998MNRAS.295..319M}).
It is, however, not yet clear whether spin parameters of galaxies and 
their host halos are tightly correlated. Hydrodynamical cosmological 
hydrodynamical simulations 
(e.g., \cite{2015MNRAS.449.2087D,2015MNRAS.450.2327Z,2020MNRAS.497.4346K}) 
suggested that spin parameters (also size) of galaxies are somewhat 
determined by initial angular momenta of their parent dark matter halos, 
but the correlation is possibly non-linear. Consistently, \cite{2021arXiv210112373D} 
showed somewhat signatures of correlation that halos having lower spin 
parameter are likely to form more compact galaxies in the IllustrisTNG 
simulations. \cite{2019MNRAS.488.4801J}, however, found a weak, if any, 
correlation using the galaxies from the NIHAO project. 
A method that can estimate the spin parameters of halos 
and protohalos sufficiently will permit us to predict the spins of 
galaxies better.
As galaxy number densities and spins can be used to reconstruct 
the initial conditions in Lagrangian space,
it is intuitive to study the Lagrangian properties of protohalos and
protogalaxies and explore how they are related
to the conservation and predictability of halo spins in Eulerian Space.

The rest of the paper is structured as follows.
In Sec.\ref{sec.mode}, we review the construction of the spin mode.
In Sec.\ref{sec.par}, we define new spin parameters in Lagrangian space 
to qualify the misalignment between moment of inertia and tidal tensor.
In Sec.\ref{sec.sim&result}, we present the results done by $N$-body simulations.
Discussions and prospects are presented in Sec.\ref{sec.fin}.

\section{Spin conservation and reconstruction}\label{sec.mode}

In this section we start with a numerical study of spin conservation
while defining some of the key parameters.
We use $j$ to denote the angular momentum of an object
while use $L$ to denote any angular momentum fields.
The Eulerian angular momentum vector of a dark matter halo is defined as
$\spin_E=\sum_p m_p (\bs{x}_p-\bar{\bs{x}})\times \bs{v}_p$,
where $m_p$, $\bs{x}_p$ and $\bs{v}_p$ are the mass, Eulerian position and
velocity of the $N$-body particles contained in a halo, and $\bar{\bs{x}}$ is 
its center of mass.
Because the Lagrangian space coordinates can be identified by using 
particle-IDs, the Lagrangian space angular momentum vector is computed
as 
\begin{eqnarray}\label{eq.spinL}
	\spin_L=\sum_p m_p\bs{q}_p'\times\bs{u}_p
=\sum_p m_p\bs{q}_p'\times(-\nabla\phi|_{\bs{q}_p}),
\end{eqnarray}
where 
$\bs q_p$, $\bar{\bs{q}}$, $\bs u_p$ denote positions, 
center of mass, velocities in Lagrangian space, and
$\bs{q}_p'\equiv\bs{q}_p-\bar{\bs{q}}$.
$\bs{u}_p$ is expressed by the gradient of the primordial 
gravitational potential $\phi$ \cite{1970A&A.....5...84Z}.
We quantify the correlation between two spin directions as
the cosine of their opening angle
$\mu(\bs{v}_1,\bs{v}_2)=\cos\theta=\hat{\bs v}_1 \cdot\hat{\bs v}_2\in[-1,1]$.
In three-dimensional space, $\mu$ of two random (uncorrelated) vectors obeys a 
top-hat distribution with $\langle\mu\rangle=0$.

In a set of LSS $N$-body simulations generated 
by {\small CUBE} \cite{2018ApJS..237...24Y},
we identify dark matter halos by the friend of friend (FoF) method with 
the linking parameter $b=0.2$ at different redshifts. 
In Fig.\ref{fig.mu_z} we plot the probability distribution functions (PDFs) 
of $\mu(\spin_L,\spin_E)$ for different halo mass bins and redshifts.
Here, the data is obtained from a flat $\Lambda$CDM simulation 
with cosmological parameters 
$\Omega_m=0.3, \Omega_\Lambda=0.7, \sigma_8=0.87, h=0.7$.
A total of $512^3$ $N$-body particles are initialized at redshift
$z_{\rm init}=100$ using the Zel'dovich approximation 
\cite{1970A&A.....5...84Z} and are evolved 
using the particle-particle particle-mesh (P3M) force calculation,
in a periodic cubic box with $L=100\,{\rm Mpc}/h$ per side.
The particle mass is thus $\simeq 8.8\times10^8 M_\odot$ 
and the least massive halos considered ($2\times 10^{11}M_\odot$) 
have $\simeq 230$ particles.
The overall PDFs of $\mu(\spin_L,\spin_E)$ obviously depart from a 
tophat distribution, suggesting that $\spin_L$, $\spin_E$ directions 
are strongly correlated.
The various PDF curves show the redshift and mass dependences.
We can see that $\mu(\spin_L,\spin_E)$ tends to be higher for a 
fixed halo mass bin but at earlier epoch of the universe.
For example, the expectation value $\langle\mu\rangle$ takes 
$0.78$, $0.71$ and $0.61$ for halos
$10^{12}<M/M_\odot<3\times10^{12}$, at redshifts
2.0, 1.0 and 0.0 respectively.
Moreover, $\mu(\spin_L,\spin_E)$ is slightly higher for more massive
halos at a fixed redshift.

To test the numerical convergence, we study $\mu(\spin_L,\spin_E)$ 
in a set of simulations with different mass and force resolutions.
Mass resolutions are modified by smaller or larger box sizes 
but fixed grid and particle numbers,
whereas force resolutions are configured by changing the
grid number to particle number ratio, particle-partile force range etc.
We find that higher mass or force resolution leads to only slightly
higher $\mu(\spin_L,\spin_E)$ for given halo mass bins.
For these reasons, we use the simulation configurations listed in
the last paragraph and selected the lower halo mass bound
to be $2\times 10^{11}M_\odot$ (230 $N$-body particles) so 
that the results in Fig.\ref{fig.mu_z} and in the rest of the paper
converge and is reliable.

In the followings we use component equations involving vectors and tensors,
and Einstein summation for repeated subscripts is implicitly assumed.
Because the Lagrangian space corresponds to the initial, homogeneous state of the
matter distribution, in Eq.(\ref{eq.spinL}) the sum over particles is 
equivalent to integration over $V_q$ -- the protohalo region which eventually 
collapses into the halo. 
Further expanding the $-\nabla\phi$ term with respect to 
$\bar{\bs{q}}$ to the first order  enables us to express 
the tidal torque prediction \cite{1984ApJ...286...38W} of $\spin_L$ as
\begin{eqnarray}\label{eq.L_IT}
	L_{IT}&=&
	\left. \int_{V_q} \epsilon_{ijk} q'_j (-\phi_{,k}) {\rm d}V \right\vert_{\text{1st-order}} \nonumber\\
			&=&\int_{V_q} \epsilon_{ijk} q'_j \left(\left.-\phi_{,kl}\right|_{\bar{\bs{q}}} q'_l\right) {\rm d}V 
			=\epsilon_{ijk} I_{jl} T_{lk},
\end{eqnarray}
where the 3D Levi-Civita symbol $\epsilon_{ijk}$
is used for the cross product in Eq.(\ref{eq.spinL}), and in the last equality
the moment of inertia tensor of $V_q$ is defined as 
$(\I)_{ij}=I_{ij}=\int_{V_q}q'_i q'_j{\rm d}V$ and the
tidal tensor is defined as $(\T)_{ij}=T_{ij}=-\phi_{,ij}$.
From Eq.(\ref{eq.L_IT}) we see that the $\Spin_{IT}$ field is a proxy of $\spin_L$.
Similar to Fig.\ref{fig.mu_z}, we also confirmed the strong correlation
in terms of $\mu(L_{IT},\spin_L)$ and $\mu(L_{IT},\spin_E)$ and
they have similar mass and redshift dependences as $\mu(\spin_L,\spin_E)$.
However, since $I_{ij}$ can not be obtained from the first principal, 
we are more interested in a predictable spin field given 
an initial density reconstruction.
In \cite{2020PhRvL.124j1302Y} we construct a spin field from a ``tide-tide''
self-interaction $L_{TT}=\epsilon_{ijk}T^r_{jl}T^R_{lk}$ where 
$r$ and $R$ are two smoothing scales smoothed on the tidal tensor. 
Choosing Gaussian smoothing kernels with $R\rightarrow r_+$, 
the spin reconstruction can be approximated by
\begin{eqnarray}\label{eq.L_TT}
	L_{TT}(r)=\epsilon_{ijk}\phi_{,jl}(r)\delta_{,lk}(r)
\end{eqnarray}
where $\delta$ is the initial overdensity field.
We have confirmed that $\Spin_{TT}$ is a reconstructed spin field as a
reliable proxy of $\spin_L$, $\spin_E$ of dark matter halos in $N$-body simulations
\cite{2020PhRvL.124j1302Y}, of $\spin_E$ of simulated galaxies in galaxy formation
simulations (Liao {\it et al.} in prep.), and of $\spin_E$ of observed disk galaxies
\cite{2020NatAs.tmp..241M}.

\section{Spin parameters in Lagrangian space}\label{sec.par}
So far we have discussed the spin correlations between $\spin_E$,
$\spin_L$, $\Spin_{IT}$ and $\Spin_{TT}$ in terms of their directions.
It is also intuitive to study the correlations between the magnitudes of these vectors.
The spin magnitudes of dark matter halos have been extensively discussed.
For example, a dimensionless spin parameter of a virialized object is defined as
$\lambda_P\equiv J|E|^{1/2}G^{-1}M^{-5/2}$ \cite{1969ApJ...155..393P}
or $\lambda_B\equiv J/(\sqrt{2}MVR)$ \cite{2001ApJ...555..240B}.
Here $J$, $E$, $M$, $V$, $R$ are the total angular momentum, total energy, 
mass, circular velocity, radius of the system, and $G$ is the Newton's constant. 
It is shown that $\lambda_P$ and $\lambda_B$ are similar for 
typical NFW halos \cite{2001ApJ...555..240B},
however not all the above quantities are straightforwardly defined for a protohalo
in the Lagrangian space. Here we introduce some new parameters to characterize the 
spin magnitudes of Lagrangian halos.
According to the kinematics of halo particles in Lagrangian space,
a dimensionless kinematic spin parameter $\lambda_K$ can be defined as
\begin{eqnarray}\label{eq.lambda_K}
	\lambda_K&\equiv&\frac
	{\int_{V_q}\hat{j_i}\epsilon_{ijk}q'_j u'_k\diff M}
	{\int_{V_q}q'u'\diff M}
	=\frac
	{\int_{V_q}\sin\theta_1\cos\theta_2 q'u'\diff M}
	{\int_{V_q}q'u'\diff M},
\end{eqnarray}
where $\hat{j_i}=(\spin_L/j_L)_i$ is the unit $\spin_L$ vector ($j_L=|\spin_L|$),
$q'=|{\bs q}'|$, $u'=|{\bs u}'|$, 
$\sin\theta_1=\sqrt{1-\mu^2({\bs q}',{\bs u}')}$ and 
$\cos\theta_2=\mu(\bs{q}'\times\bs{u}',\spin_L)$.
Clearly, $\lambda_K\in[0,1]$ characterizes the rotation-supportedness 
of the system.
A coplanar system with all mass elements having homodromous circular orbits,
$\lambda_K=1$, while a rotating rigid isodensity globe has $\lambda_K=8/3\pi\simeq 0.85$.

Similar to the fact that $\Spin_{IT}$ and $\Spin_{TT}$ being 
the proxies of $\spin_L$ by the tidal torque theory, 
we can also find proxies of $\lambda_K$ in Lagrangian space.
We expect that the rotation-supportedness $\lambda_K$ of a Lagrangian system depends on\\
(a) the anisotropies of the tidal field,
(b) the nonspherical distribution of the protohalo, and
(c) the misalignment between the principal axes of the above two.
Both (a) and (b) can be characterized by the 
elipticity $e=(\lambda_1-\lambda_3)/2(\lambda_1+\lambda_2+\lambda_3)$ and
prolateness $p=(\lambda_1-2\lambda_2+\lambda_3)/2(\lambda_1+\lambda_2+\lambda_3)$
of the tensors $\I$ and $\T$, where $\lambda_{1,2,3}$ are the
primary, intermediate and minor eigenvalues of the tensor \cite{2002MNRAS.332..339P}.
For (c) \cite{2009ApJ...707..761L} has defined an alignment parameter 
$\beta\equiv 1-\sqrt{
(\varrho^2_{12}+\varrho^2_{23}+\varrho^2_{31})/
(\varrho^2_{11}+\varrho^2_{22}+\varrho^2_{33})}$,
where $\varrho_{ij}$'s are the re-expressions of elements of $\I$ in the
principal frame of $\T$.
We find weak correlations between $\lambda_K$ and these parameters.
Here we define dimensionless parameters
\begin{eqnarray}
	\beta_{IT}&\equiv& \left. 
	\frac{\left\vert\int_{V_q}\epsilon_{ijk}q'_{j}u'_{k}\diff V\right\vert}
	{-\int_{V_q}q'_i u'_i\diff V}\right\vert_{\text{1st-order}} \nonumber\\
	&=&
	\frac{\left\vert\int_{V_q} 
	\epsilon_{ijk}q'_j(-\phi_{,kl}q'_l)\diff V\right\vert}
	{-\int_{V_q}q'_i(-\phi_{,ij}q'_j)\diff V}
	= \frac{\left\vert\epsilon_{ijk}I_{jl}T_{lk}\right\vert}{-I_{ij}T_{ij}}
\end{eqnarray}
and similarly
\begin{eqnarray}
	\beta_{TT}(r)&\equiv&
	= \frac{\left\vert\epsilon_{ijk}\phi_{,jl}(r)\delta_{,lk}(r)\right\vert}
	{-\phi_{,ij}(r)\delta_{,ij}(r)}
\end{eqnarray}
as the proxies of $\lambda_K$,
where $\beta_{IT}$ optimizes the measurement of anisotropies and misalignment
between two tensors $\I$ and $\T$, 
whereas tidal environment parameter $\beta_{TT}$ 
measures the anisotropy and twist of $\T$ on physical scale $r$.
A schematic interpretation of $\beta_{TT}$ is discussed in the figure 1 of
\cite{2020NatAs.tmp..241M}.

In the spin mode reconstruction \cite{2020PhRvL.124j1302Y}, 
in order to optimize the correlation $\mu(\spin_L,\spin_E)$,
the smoothing scale $r$ is chosen to match the equivalent protohalo scale 
$r=r_q(M)=(2MG/\Omega_m H_0^2)^{1/3}$, otherwise $\spin_L$ and $\spin_E$ are decorrelated.
Here $\Omega_m$ is the density parameter for matter and $H_0$
is the Hubble constant.

\section{Spin Conservation and Predictability}\label{sec.sim&result}

\begin{figure}
	\centering
		\includegraphics[width=1\columnwidth]{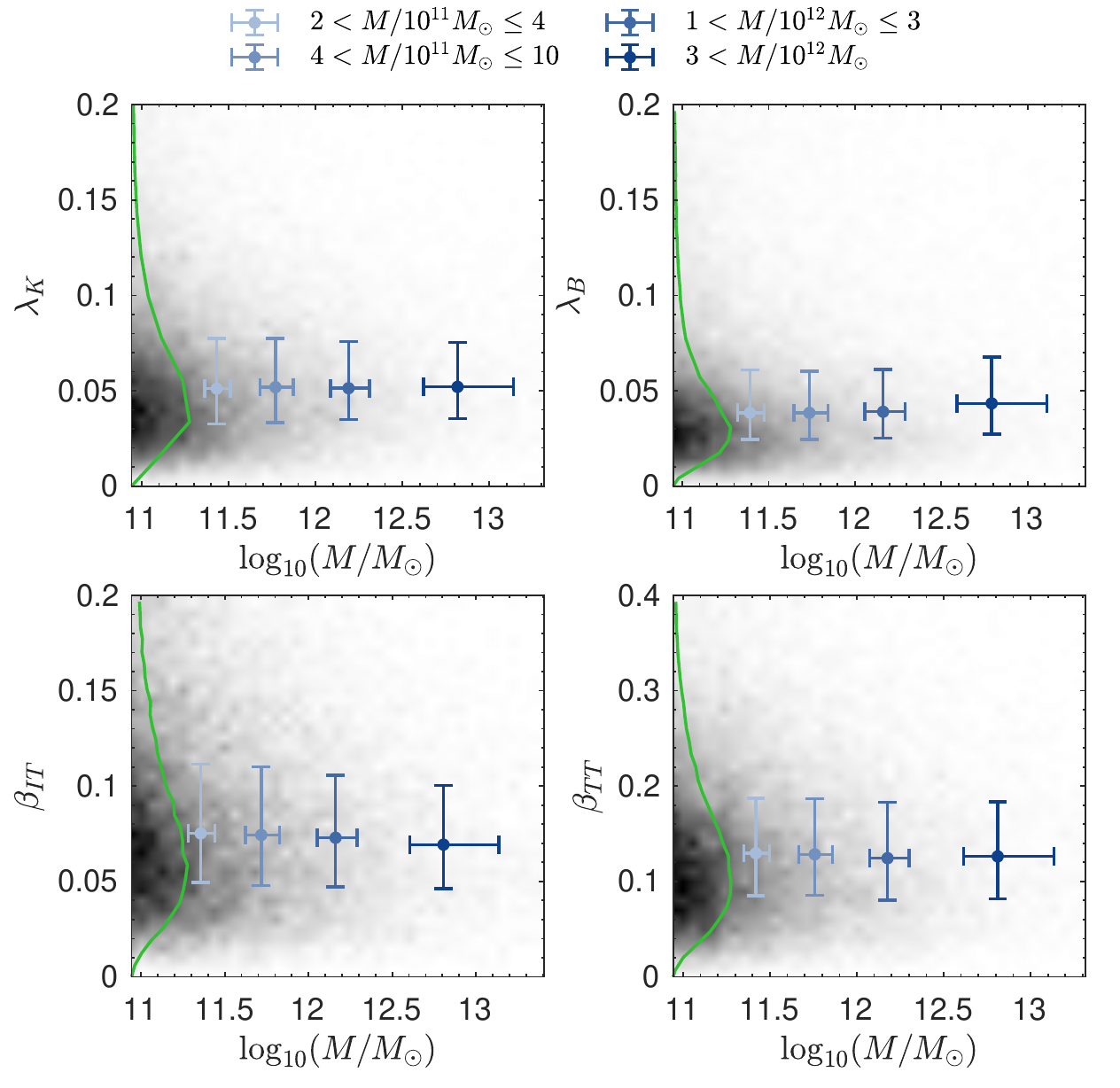}
	\caption{PDFs of $\lambda_K$, $\lambda_B$, $\beta_{\rm IT}$, $\beta_{\rm TT}$
	and their halo mass dependences. 
	The four green curves show the PDFs of these parameters and the gray scale maps 
	represent the halo number density in the two-dimensional parameter space.
	The normalizations are arbitrarily set for a clear visualization.
	We partition all halos into four mass groups specified on the top of the
	figure, and inside each mass group, we take the three quartiles $\{Q_1,Q_2,Q_3\}$
	of the parameter distribution as the lower/left boundary of the error bar, 
	center of the error bar, and upper/right boundary of the error bar.}
	\label{fig.params_vs_mass}
\end{figure}

The following results are obtained from the $z=0$ checkpoint 
of a simulation described in Sec.\ref{sec.mode}.
The numerical convergence is also verified by using a set of
companion simulations with different mass and force resolutions as mentioned in
Sec.\ref{sec.mode}

In Fig.\ref{fig.params_vs_mass} we plot the PDFs of $\lambda_K$, 
$\lambda_B$, $\beta_{\rm IT}$, $\beta_{\rm TT}$ and their
joint PDFs with halo mass.
The green curves in the four panels show the PDFs of 
the indicated parameters for all halos with at least 100 particles 
($>8.8\times 10^{10}M_\odot$), however we note that they are insensitive
for selected halo mass bins.
The Lagrangian space spin parameter $\lambda_K$ has a very similar
distribution as $\lambda_B$. 
In particular, $\langle\lambda_K\rangle\simeq 0.063$
with a standard deviation $\sigma\simeq 0.041$,
in comparison, $\langle\lambda_B\rangle\simeq 0.051$ with a standard deviation 
$\sigma\simeq 0.037$.
All these parameters show a lognormal probability distribution,
with the expectation value $\leq 0.1$.
We use gray-scale color maps to show the two-dimensional PDFs
of each parameter and halo mass.
To clearly illustrate the mass dependences here and for the following
studies, we select all halos which are more massive than
$2\times 10^{11}M_\odot$, and partition them into four mass bins.
Note that, due to the behaviour of the halo mass function, 
a significant portion of less massive halos are beyond the lower bound
of the lowest mass bin, however whether including them does not change
the statistics.
For each mass bin, we plot the center of the error bar at the median
of halo mass and the median of the parameter in consideration.
The lower/left and upper/right boundary of each error bar are taken as the
25\% and 75\% percentiles of the distribution.
Similar to that of $\lambda_B$, 
we do not see obvious mass dependences of other parameters
in Fig.\ref{fig.params_vs_mass}.

\begin{figure}
	\centering
		\includegraphics[width=1\linewidth]{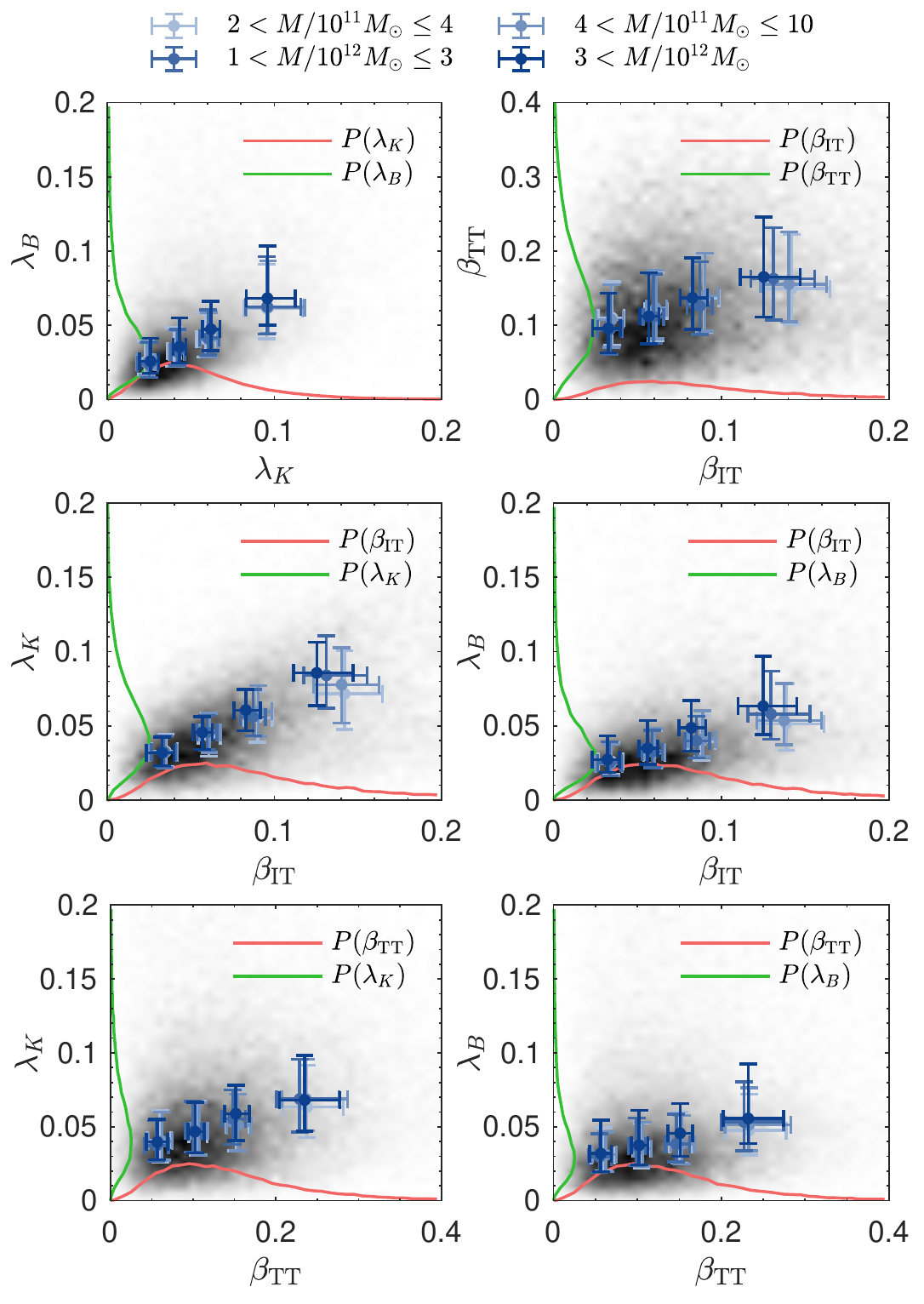}
	\caption{Dependences between $\lambda_K$, $\lambda_B$, $\beta_{\rm IT}$ and $\beta_{\rm TT}$.
	The gray scale in the background represents their distribution in 
	two-dimensional parameter space, and halos are divided into four mass bins
	specified on the top of the figure.
	The red and green curves represent the PDFs of the parameters indicated by the 
	label of each axis.
	The data are partitioned into four mass bins as shown on the top of the figure,
	and then are sorted by the value of parameters in the $x$-axis,
	with the positions and boundaries the same as Fig.\ref{fig.params_vs_mass}. }
	\label{fig.CCR_of_Params}
\end{figure}

In Fig.\ref{fig.CCR_of_Params}, we plot the dependences between
$\lambda_K$, $\lambda_B$, $\beta_{IT}$ and $\beta_{TT}$ in six panels.
For each pair of parameters we use the gray scale to plot the halo number densities 
in their two-dimensional parameter space,
and the two curves show their individual PDF.
Again, all halos are partitioned into four mass bins indicated,
and the positions and boundaries of the error bars are determined
similarly as Fig.\ref{fig.params_vs_mass}.
A subtle difference is that, for each mass bin,
the data are subsequently sorted according to the parameter in
the horizontal axis ($\lambda_K$ in the top-left panel, for example)
and then partitioned into four groups of equal (may differ up to one) number of halos.
Then, for each group, the error bar position and boundaries are determined
by the two parameters ($\lambda_K$ and $\lambda_B$ in the top-left panel, for example)
in consideration. Thus there are 16 error bars in each panel, and we
use gray scales of the error bar to indicate mass, and for each
mass bin, the four groups are naturally placed horizontally.
For all panels, we see positive correlations between all six possible combinations
of four parameters despite of the certain scatters.
Firstly, in the top-left panel, either by the 2D PDF or the error bars,
we see a strong positive correlation between $\lambda_K$ and $\lambda_B$,
and the correlation appears in all halo mass bins.
It shows that the kinematic spin-supportedness of Lagrangian halos is closely
related to the Eulerian spin parameter --  a more spin supported Lagrangian
protohalo is more likely to result in a spin supported halo at $z=0$.
In the middle-left panel, the correlation between $\beta_{IT}$ and
$\lambda_K$ shows that a more isotropic $\I$ and $\T$ and a more misalignment
between the two indeed lead to a more spin-supported Lagrangian protohalo.
The bottom-left and top-right panels show that $\beta_{TT}$ is indeed a
proxy of $\beta_{IT}$ and $\lambda_K$, making the prediction of Lagrangian
spin parameter applicable.
Finally, the middle-right and bottom-right panels also illustrate
positive correlations between $\beta_{IT}$ - $\lambda_B$ and 
$\beta_{TT}$ - $\lambda_B$ respectively, although the correlations
are weaker compared to the correlations measured directly in Lagrangian space.

\begin{figure}
	\centering
		\includegraphics[width=\columnwidth]{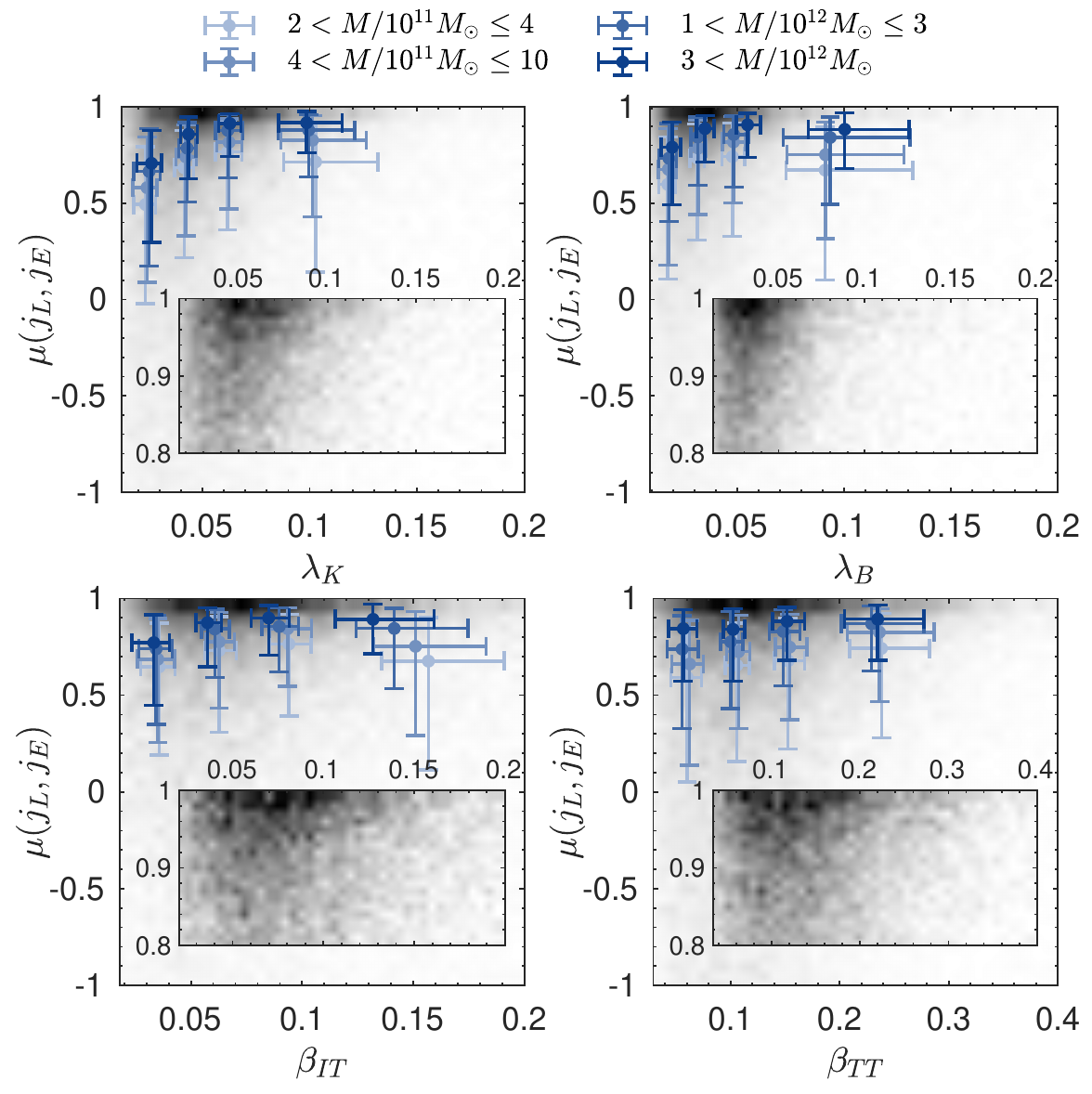}
	\caption{PDFs of $\lambda_K$, $\lambda_B$, $\beta_{\rm IT}$, $\beta_{\rm TT}$
	and the spin conservation quantified by $\mu(\spin_L,\spin_E)$.
	The gray-scale color maps show the two-dimensional PDFs of each parameter and
	correlation function $\mu(\spin_L,\spin_E)$.
	The bottom right inset of each panel is the zoom-in PDF of 
	the most dense part.
	The data are partitioned into four mass bins and sorted the same as 
	Fig.\ref{fig.CCR_of_Params}.}
	\label{fig.params_mu_le}
\end{figure}

Next we investigate how the spin conservation depends on 
the spin parameters and Lagrangian properties.
In Fig.\ref{fig.params_mu_le} we choose the horizontal axes as
the $\lambda_K$, $\lambda_B$, $\beta_{\rm IT}$ and $\beta_{\rm TT}$ and the
vertical axes as $\mu(\spin_L,\spin_E)$ and use the gray-scale maps to show 
the halo number densities on these parameter spaces.
The error bars are plotted the same as in Fig.\ref{fig.CCR_of_Params}.
In the top-left panel, we clearly see that, in Lagrangian space, 
and for all the four mass bins,
spin-supported (high-$\lambda_K$) protohalos tend to preserve their
spin directions all the way to $z=0$, leading to a systematically
higher $\mu(\spin_L,\spin_E)$.
For these halos, the spins observed in Eulerian space contain more information
if we directly use the Eulerian spin to study the primordial universe.
In the top-right panel we see the same trend when the horizontal axis is
replaced by $\lambda_B$, except that the highest $\lambda_B$ shows a 
slight decrease in $\mu(\spin_L,\spin_E)$ for all mass bins.
In the bottom-left panel, $\mu(\spin_L,\spin_E)$ is also positively
correlated with $\beta_{\rm IT}$, except for the three lowest mass bins,
where the highest $\beta_{\rm IT}$ leads $\mu(\spin_L,\spin_E)$ to decrease.
Most importantly, in the bottom-right panel, we see a consistent positive 
correlation between $\mu(\spin_L,\spin_E)$ and $\beta_{\rm TT}$ for all
mass bins. 
$\beta_{\rm TT}$ can be directly calculated by the reconstructed initial
density field, so in principle we can weight those halo-galaxy systems
with high $\beta_{\rm TT}$ values such that their observed spins better reflect
cosmological information in Lagrangian space.

\begin{figure}
	\centering
		\includegraphics[width=1\columnwidth]{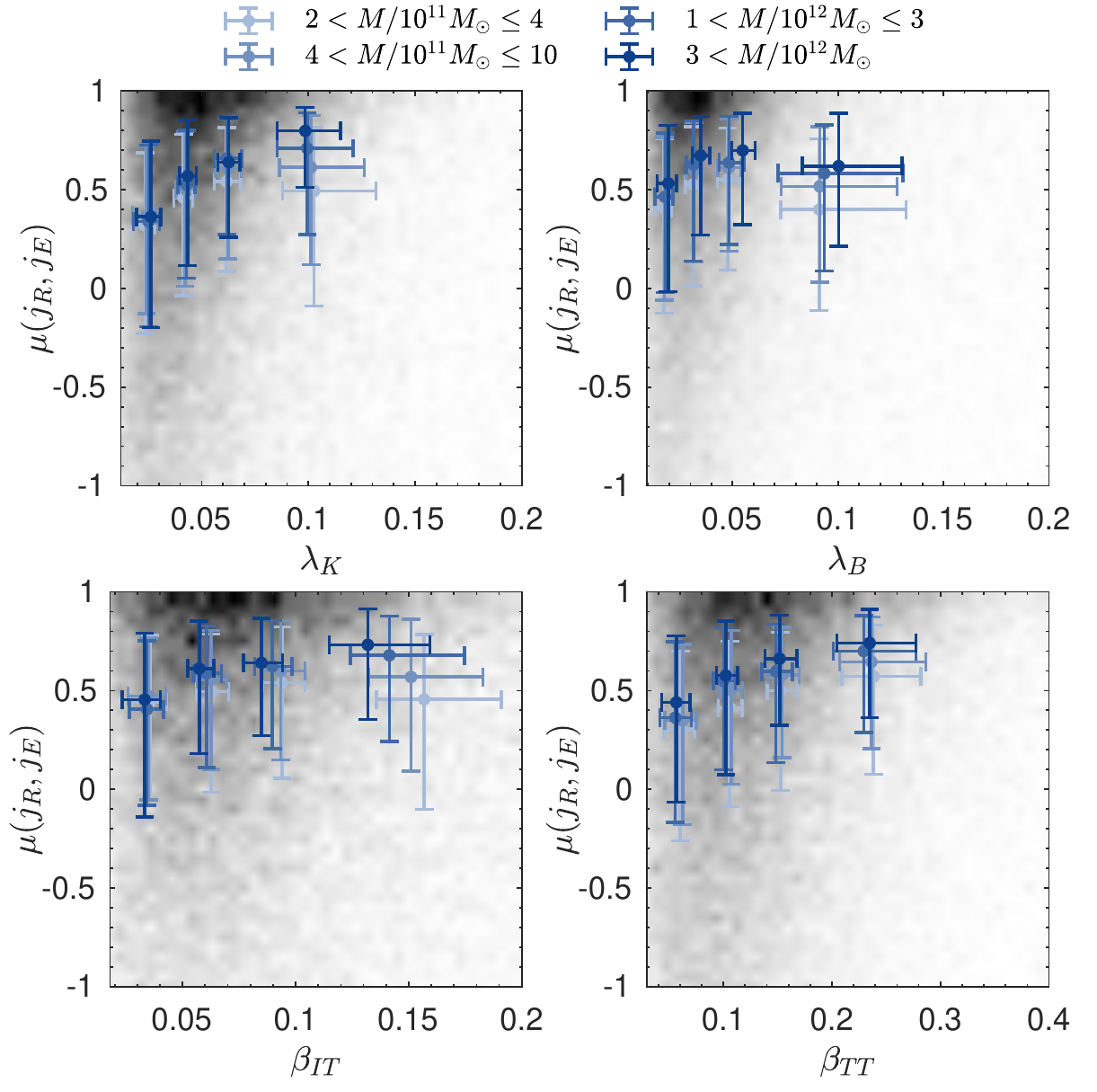}
	\caption{PDFs of $\lambda_K$, $\lambda_B$, $\beta_{\rm IT}$, $\beta_{\rm TT}$
	and the spin conservation quantified by $\mu(\spin_R,\spin_E)$.
	The gray-scale color maps and error bars are the same as in Fig.\ref{fig.params_mu_le},
	except the horizontal axes are set as $\mu(\spin_R,\spin_E)$. }
	\label{fig.params_mu_re}
\end{figure}

Next, we reconstruct the predicted halo spins by the Eq.(\ref{eq.L_TT})
with the optimal smoothing scale 
$r=r_{\rm opt}=(2M_{\rm bin}G/\Omega_m H_0^2)^{1/3}$,
where $M_{\rm bin}$ is the averaged halo mass of the corresponding mass
bin. In \cite{2020PhRvL.124j1302Y} we have demonstrated that a rough estimate
of the halo mass can give accurate spin reconstruction, not sensitive
to the mass estimation.
The predicted halo spin is thus $\spin_R=\Spin_{TT}(r_{\rm opt})$.
The predictability of halo spins is quantified by $\mu(\spin_R,\spin_E)$.
In Fig.\ref{fig.params_mu_re} we plot $\mu(\spin_R,\spin_E)$ as a function
of parameters $\lambda_K$, $\lambda_B$, $\beta_{\rm IT}$ and $\beta_{\rm TT}$
similar to Fig.\ref{fig.params_mu_le}.
Compared to Fig.\ref{fig.params_mu_le}, the dependences of $\mu(\spin_R,\spin_E)$
to $\lambda_K$, $\lambda_B$, $\beta_{\rm IT}$, $\beta_{\rm TT}$ 
show similar trends to that of $\mu(\spin_L,\spin_E)$ 
but with more dispersion in the distribution of spin predictability.
Importantly, in the bottom right panel, 
we see positive correlation between $\beta_{TT}$ and 
$\mu(\spin_R,\spin_E)$ for all mass bins. 
From density reconstruction, we can calculate both $\beta_{TT}$ and 
$\spin_R$ of each halo, and this correlation suggests that halos with
higher $\beta_{TT}$ values tend to have more reliable predicted
halo spins $\spin_R$. So we can directly estimate the accuracy of 
spin reconstruction for all halos, and improve the accuracy of the
reconstructed initial condition.

\section{Conclusion}\label{sec.fin}
We study the conservation and predictability of the angular momenta of 
dark matter halos depending on their spin parameters 
and Lagrangian space properties.

Similar to the definitions of halo spin parameters in Eulerian space,
we construct a dimensionless kinematic spin parameter $\lambda_K$ 
of protohalos in Lagrangian space. 
Higher $\lambda_K$ corresponds to more rotational-supported systems.
According to the tidal torque theory, we introduce the parameter
$\beta_{\rm IT}$ to characterize the anisotropies of $\I$ and $\T$ 
and the misalignment between the two, where $\I$ is the moment of 
inertia tensor of the protohalo and $\T$ is the tidal tensor it feels.
The kinematics of a low $\beta_{\rm IT}$ protohalo is more dominated
by a radial convergence flow rather than coherent circular motions.
Our simulations show that these low $\beta_{\rm IT}$ protohalos 
typically have low $\lambda_K$ and are eventually more likely to
evolve into velocity dispersion supported (low $\lambda_B$) halos.
Similar to the spin mode reconstruction in \cite{2020PhRvL.124j1302Y}, 
we also reconstruct the tidal torque scalar $\beta_{\rm TT}$ from 
initial density fields and it turns out that $\beta_{\rm TT}$
is a reliable proxy of $\beta_{\rm IT}$ to characterize 
the spin parameters.

We quantify the conservation of halo spins as the correlation between their
Lagrangian and Eulerian spin directions $\mu(\spin_L,\spin_E)$.
From simulations we see that rotation supported protohalos 
(higher $\lambda_K$, and thus statistically higher
$\lambda_B$, $\beta_{\rm IT}$ and $\beta_{\rm TT}$)
tend to better preserve their spin during the cosmic evolution, 
and result in a higher $\mu(\spin_L,\spin_E)$.
This is understandable -- for less rotation supported systems
it is difficult to determine their total angular momentum vectors
and the rather random $\spin_L$ and $\spin_E$ result in a random
correlation between the two.
We also show that the spin predictability by Eq.(\ref{eq.L_TT})
\cite{2020PhRvL.124j1302Y} is positively correlated with these spin parameters.

The spins of disk galaxies are observable and we have discovered a 
positive correlation signal between disk galaxies and their halos 
and the reconstructed 
initial density field of the local universe \cite{2020NatAs.tmp..241M}.
This observed correlation is a result of many facts:
the observability of galaxy spiral arms indicating their spin
directions; the underlying galaxy-halo spin correlation, confirmed
by many galaxy formation simulations; the Lagrangian-Eulerian
spin correlation; the tidal torque theory and spin mode reconstruction
linking the primordial density field with protohalo spins;
and the density reconstruction given galaxy positions.
Galaxy spins are proposed to discover properties of the primordial
universe such as neutrino mass, chiral violations, primordial 
non-Gaussianities etc, and a better understanding of the above 
facts and improve the correlations will be the first step.
This paper shows that Lagrangian spin parameter $\lambda_K$ 
is closely related to Eulerian ones $\lambda_B$, 
and to the Lagrangian tidal properties, and is 
predictable.
Therefore, a weighted selection of rotation-supported halos 
by $\beta_{TT}$ is expected to straightforwardly improve 
the spin correlation to the	primordial universe.

The spin reconstruction algorithms consider dark matter halos and protohalos and
do not include any baryonic effects.
This is because we have used the Lagrangian space information, i.e.,
the primordial tidal environment properties, to predict the spin
in the Eulerian space.
The baryonic and dark matter components of a galaxy-halo system have very similar 
protovolumes in Lagrangian space \cite{2017MNRAS.470.2262L}, 
so they feel the same direction and magnitude of the initial tidal torque, 
which is constributed by the initial total gravity.
This is the reason why the observation of galaxy spins are correlated with
the dark matter (or total) initial conditions \cite{2020NatAs.tmp..241M}.
This work uses dark matter only simulations and focuses on how spin parameters
influence the spin conservation and predictability of dark matter halos.
Since galaxy and halo spin directions are tightly 
correlated \cite{2015ApJ...812...29T},
better predictability of halo spin directions indicates better predictability of galaxy spin
directions, so the spin parameters of protohalos are also related to the
predictability of galaxy spin directions.
Our method can also be applied to investigate the correlation 
of spin magnitudes between galaxies and their host halos, 
which is out of the scope of this paper.

The kinematic spin parameter can also be used to describe the rotation
of cosmic filaments. The filament spins are confirmed
by observations of the peculiar velocities of 
galaxies \cite{2020ApJ...900..129W} and by numerical
simulations \cite{2020arXiv200602418X}.
In Lagrangian space, the origin of the filament spin could also be
described by the initial tidal field.
On the other hand, the Eulerian filaments are non-virialized 
systems so we expect their spins correlated with Lagrangian ones.
In both Eulerian and Lagrangian spaces, the spin magnitudes can
be quantified by $\lambda_K$.
Additionally, in cosmic filaments, it would be interesting to study 
the correlations between large scale spin mode by galaxy peculiar
velocities and small scale spin modes by galaxy spins.

\section*{Acknowledgments}
We thank Pavel Motloch and Ue-Li Pen for valuable discussions and comments.
We thank the anonymous referee for valuable suggestions.
H.R.Y. is supported by NSFC 11903021.
S.L. acknowledges the support by the European Research Council via 
ERC Consolidator grant KETJU (No. 818930).
M.D. is supported by the grants 
``National Postdoctoral Program for Innovative Talents'' 
(\#8201400810) and 
``Postdoctoral Science Foundation of China'' 
(\#8201400927).
The simulations were performed on the workstation of cosmological sciences,
DoA, XMU.

\bibliography{haoran_ref}

\begin{thebibliography}{10}

\bibitem{1969ApJ...155..393P}
P.~J.~E. {Peebles},
\newblock \apj {\bf 155}, 393 (1969).

\bibitem{1970Afz.....6..581D}
A.~G. {Doroshkevich},
\newblock Astrofizika {\bf 6}, 581 (1970).

\bibitem{1984ApJ...286...38W}
S.~D.~M. {White},
\newblock \apj {\bf 286}, 38 (1984).

\bibitem{2000ApJ...532L...5L}
J.~{Lee} and U.-L. {Pen},
\newblock \apj {\bf 532}, L5 (2000), astro-ph/9911328.

\bibitem{2001ApJ...555..106L}
J.~{Lee} and U.-L. {Pen},
\newblock \apj {\bf 555}, 106 (2001), astro-ph/0008135.

\bibitem{2015ApJ...812...29T}
A.~F. {Teklu} {\em et~al.},
\newblock \apj {\bf 812}, 29 (2015), 1503.03501.

\bibitem{2017PhRvD..95d3501Y}
H.-R. {Yu}, U.-L. {Pen}, and H.-M. {Zhu},
\newblock \prd {\bf 95}, 043501 (2017), 1610.07112.

\bibitem{2020PhRvL.124j1302Y}
H.-R. {Yu} {\em et~al.},
\newblock \prl {\bf 124}, 101302 (2020), 1904.01029.

\bibitem{2016ApJ...831..164W}
H.~{Wang} {\em et~al.},
\newblock \apj {\bf 831}, 164 (2016), 1608.01763.

\bibitem{2020NatAs.tmp..241M}
P.~{Motloch}, H.-R. {Yu}, U.-L. {Pen}, and Y.~{Xie},
\newblock Nature Astronomy  (2020), 2003.04800.

\bibitem{2019PhRvD..99l3532Y}
H.-R. {Yu}, U.-L. {Pen}, and X.~{Wang},
\newblock \prd {\bf 99}, 123532 (2019), 1810.11784.

\bibitem{1983IAUS..100..391F}
S.~M. {Fall},
\newblock {Galaxy formation - Some comparisons between theory and observation},
\newblock in {\em Internal Kinematics and Dynamics of Galaxies}, edited by
  E.~{Athanassoula}, volume 100, pp. 391--398, 1983.

\bibitem{2012ApJS..203...17R}
A.~J. {Romanowsky} and S.~M. {Fall},
\newblock \apjs {\bf 203}, 17 (2012), 1207.4189.

\bibitem{1980MNRAS.193..189F}
S.~M. {Fall} and G.~{Efstathiou},
\newblock \mnras {\bf 193}, 189 (1980).

\bibitem{1998MNRAS.295..319M}
H.~J. {Mo}, S.~{Mao}, and S.~D.~M. {White},
\newblock \mnras {\bf 295}, 319 (1998), astro-ph/9707093.

\bibitem{2015MNRAS.449.2087D}
M.~{Danovich}, A.~{Dekel}, O.~{Hahn}, D.~{Ceverino}, and J.~{Primack},
\newblock \mnras {\bf 449}, 2087 (2015), 1407.7129.

\bibitem{2015MNRAS.450.2327Z}
A.~{Zolotov} {\em et~al.},
\newblock \mnras {\bf 450}, 2327 (2015), 1412.4783.

\bibitem{2020MNRAS.497.4346K}
M.~{Kretschmer}, O.~{Agertz}, and R.~{Teyssier},
\newblock \mnras {\bf 497}, 4346 (2020), 2003.03368.

\bibitem{2021arXiv210112373D}
M.~{Du} {\em et~al.},
\newblock arXiv e-prints , arXiv:2101.12373 (2021), 2101.12373.

\bibitem{2019MNRAS.488.4801J}
F.~{Jiang} {\em et~al.},
\newblock \mnras {\bf 488}, 4801 (2019), 1804.07306.

\bibitem{1970A&A.....5...84Z}
Y.~B. {Zel'dovich},
\newblock \aap {\bf 5}, 84 (1970).

\bibitem{2018ApJS..237...24Y}
H.-R. {Yu}, U.-L. {Pen}, and X.~{Wang},
\newblock The Astrophysical Journal Supplement Series {\bf 237}, 24 (2018).

\bibitem{2001ApJ...555..240B}
J.~S. {Bullock} {\em et~al.},
\newblock \apj {\bf 555}, 240 (2001), astro-ph/0011001.

\bibitem{2002MNRAS.332..339P}
C.~{Porciani}, A.~{Dekel}, and Y.~{Hoffman},
\newblock \mnras {\bf 332}, 339 (2002), astro-ph/0105165.

\bibitem{2009ApJ...707..761L}
J.~{Lee}, O.~{Hahn}, and C.~{Porciani},
\newblock \apj {\bf 707}, 761 (2009), 0906.5166.

\bibitem{2017MNRAS.470.2262L}
S.~{Liao}, L.~{Gao}, C.~S. {Frenk}, Q.~{Guo}, and J.~{Wang},
\newblock \mnras {\bf 470}, 2262 (2017), 1610.07592.

\bibitem{2020ApJ...900..129W}
P.~{Wang} {\em et~al.},
\newblock \apj {\bf 900}, 129 (2020), 2007.08345.

\bibitem{2020arXiv200602418X}
Q.~{Xia}, M.~C. {Neyrinck}, Y.-C. {Cai}, and M.~A. {Arag{\'o}n-Calvo},
\newblock arXiv e-prints , arXiv:2006.02418 (2020), 2006.02418.

\end{thebibliography}

\end{document}